\documentstyle[11pt]{article}

\newcommand{\CMP}[1]{{\em Commun. Math. Phys.} {\bf {#1}}}

\newcommand{\NP}[1]{{\em Nucl.Phys.~B} {\bf {#1}}}

\newcommand{\PL}[1]{{\em Phys. Lett.} {\bf {#1}}}
\newcommand{\PR}[2]{{\em Phys. Rev.} {#1} {\bf {#2}}}
\newcommand{\PRL}[1]{{\em Phys. Rev. Lett.} {\bf {#1}}}
\newcommand{\LMP}[1]{{\em Lett. Math. Phys.} {\bf {#1}}}

\newcommand{\Map}{C_0^\infty}

\newcommand{\llim}{\lim_{\Lam\to\infty}}
\newcommand{\dvp}{d\!\!^-p}
\newcommand{\dvx}{dx}
\newcommand{\dvq}{d\!\!^-q}
\newcommand{\dvQ}{d\!\!^-Q}
\newcommand{\dvy}{dy}
\newcommand{\dvk}{d\!\!^-k}
\newcommand{\intS}{\int_{\R^d}}
\newcommand{\intF}{\int_{B^d_\Lam}}
\newcommand{\half}{\mbox{$\frac{1}{2}$}}

\newcommand{\SU}{{\rm SU}}

\newcommand{\gl}{{\rm gl}}
\newcommand{\ul}{\underline}
\newcommand{\wt}{\widetilde}

\newcommand{\eq}{\begin{equation}}
\newcommand{\eqend}{\end{equation}}
\newcommand{\eqa}{\begin{eqnarray}}
\newcommand{\nonueqa}{\begin{eqnarray*}}
\newcommand{\eqaend}{\end{eqnarray}}
\newcommand{\nonueqaend}{\end{eqnarray*}}
\newcommand{\nonu}{\nonumber \\ \nopagebreak}
\newcommand{\bma}[1]{\begin{array}{#1}}
\newcommand{\ema}{\end{array}}
\newcommand{\bc}{\begin{center}}
\newcommand{\ec}{\end{center}}

\newcommand{\Ref}[1]{(\ref{#1})}

\newcommand{\ee}[1]{\,\mbox{{\rm e}}^{#1}}
\newcommand{\ii}{{\rm i}}
\newcommand{\OO}{{\rm O}}

\renewcommand{\phi}{\varphi}
\newcommand{\sig}{\sigma}

\newcommand{\Om}{\Omega}

\newcommand{\eps}{\varepsilon}
\newcommand{\Lam}{\Lambda}
\newcommand{\sign}{{\rm sign}}

\newfam\msbfam
\batchmode\font\twelvemsb=msbm10 scaled\magstep1 \errorstopmode
\ifx\twelvemsb\nullfont\def\Bbb{\bf}
\else	\catcode`\@=11
	\font\tenmsb=msbm10 \font\sevenmsb=msbm7 \font\fivemsb=msbm5
	\textfont\msbfam=\tenmsb
	\scriptfont\msbfam=\sevenmsb \scriptscriptfont\msbfam=\fivemsb
	\def\Bbb{\relax\ifmmode\expandafter\Bbb@\else
 		\expandafter\nonmatherr@\expandafter\Bbb\fi}
	\def\Bbb@#1{{\Bbb@@{#1}}}
	\def\Bbb@@#1{\fam\msbfam\relax#1}
	\catcode`\@=\active
\fi
\newcommand{\R}{{\Bbb R}}
\newcommand{\C}{{\Bbb C}}

\newcommand{\N}{{\Bbb N}}

\newcommand{\f}{\frac}
\newcommand{\cA}{{\cal A}}

\newcommand{\cH}{{\cal H}}

\newcommand{\ccr}[2]{{[} {#1},{#2} {]} }        

\newcommand{\Tra}[1]{{\rm Tr} \left({#1}\right)}          
\newcommand{\TraL}[1]{{\rm Tr}_\Lam \left({#1}\right)}          
\newcommand{\TraC}[1]{{\rm Tr}_C \left({#1}\right)}          
\newcommand{\tra}[1]{{\rm tr} \left({#1}\right)}          
\newcommand{\tras}[1]{{\rm tr}_{\nu} \left({#1}\right)}          
\newcommand{\trac}[1]{{\rm tr}_{N} \left({#1}\right)}          
\newcommand{\BO}{B(\cH)}
\newcommand{\TR}{B_1(\cH)}

\newcommand{\SIq}{B_q(\cH)}
\newcommand{\TRC}{B_{1,C}(\cH)}

\newcommand{\g}{\ul{g}}

\newcommand{\dd}{{\rm d}}
\newcommand{\D}{D\!\!\!\!\slash}
\newcommand{\ddd}{\hat{\rm d}}
\newcounter{saveeqn}
\newcounter{App} 
\newcommand{\app}{%
\stepcounter{App}%
\setcounter{saveeqn}{\value{equation}}%
\setcounter{equation}{0}%
\renewcommand{\theequation}{\Alph{App}\arabic{equation}} }
\newcommand{\appende}{%
\setcounter{equation}{\value{saveeqn}}%
\renewcommand{\theequation}{\arabic{equation}}  }
\newcommand{\alpheqn}{%
\stepcounter{equation}
\setcounter{saveeqn}{\value{equation}}%
\setcounter{equation}{0}%
\renewcommand{\theequation}{\arabic{saveeqn}\alph{equation}} }
\newcommand{\reseteqn}{\setcounter{equation}{\value{saveeqn}}%
\renewcommand{\theequation}{\arabic{equation}} }
\newcommand{\gamd}{\gamma^{(d)}}
\newcommand{\gazd}{\gamma^{(d+2)}}
\newcommand{\slD}{\!\!\!\,\slash}
\newcommand{\sQ}{Q\!\!\slD}

\begin{document}
\pagestyle{empty}
\begin{flushright}
\today
\end{flushright}
\vspace{.4cm}
\renewcommand{\thefootnote}{\alph{footnote}}

\begin{center}
{\Large \bf Non--commutative Integration Calculus}\\
\vspace{1 cm}
{\large Edwin Langmann}\\
\vspace{0.3 cm}
{\em Theoretical Physics, Royal Institute of Technology, S-10044 Sweden}
\end{center}

\setcounter{footnote}{0}
\renewcommand{\thefootnote}{\arabic{footnote}}

\begin{abstract}
We discuss a non--commutative integration calculus arising in the
mathematical description of Schwinger terms of fermion--Yang--Mills
systems.  We consider the differential complex of forms
$u_0\ccr{\eps}{u_1}\cdots\ccr{\eps}{u_n}$ with $\eps$ a grading operator on
a Hilbert space $\cH$ and $u_i$ bounded operators on $\cH$ which naturally
contains the compactly supported de Rham forms on $\R^d$ (i.e.\ $\eps$ is
the sign of the free Dirac operator on $\R^d$ and $\cH$ a $L^2$--space on
$\R^d$).  We present an elementary proof that the integral of $d$--forms
$\int_{\R^d}\trac{X_0\dd X_1\cdots \dd X_d}$ for $X_i\in\Map(\R^d;\gl_N)$,
is equal, up to a constant, to the conditional Hilbert space trace of
$\Gamma X_0\ccr{\eps}{X_1}\cdots\ccr{\eps}{X_d}$ where $\Gamma=1$ for $d$
odd and $\Gamma=\gamma_{d+1}$ (`$\gamma_5$--matrix') a spin matrix
anticommuting with $\eps$ for $d$ even.  This result provides a natural
generalization of integration of de Rham forms to the setting of Connes'
non--commutative geometry which involves the ordinary Hilbert space trace
rather than the Dixmier trace.
\end{abstract}

\pagestyle{plain}
\setcounter{page}{1}

\newcommand{\map}{Map}

\section{Introduction}

One reason for increasing interest of physicists in non--commutative
geometry (NCG) \cite{Connes} is the hope that it could provide new powerful
tools for investigating quantum field theory beyond perturbation theory.

One example strongly supporting this hope is the representation theory of
groups $\map(M^d;G)$ of maps from an odd--$d$ dimensional Riemannian
manifold $M^d$ with spin structure to some compact semi--simple Lie group
$G$ (e.g.\ $G=\SU(N)$ in the fundamental representation) and their Lie
algebras $\map(M^d;g)$ ($g$ the Lie algebra of $G$) which are closely
related to gauge theories.  Indeed, $\map(M^d;G)$ can be regarded as the
gauge group of Yang--Mills theory on $(d+1)$--dimensional space--time
$M^d\times \R$ with structure group $G$ in the Hamiltonian framework.  One
therefore expects that quantum gauge theories should give rise to
non--trivial representations of these groups and Lie algebras satisfying
essential physical requirements such as existence of a highest weight
vector (ground state).

Both, $\map(M^d;G)$ and $\map(M^d;g)$ are subalgebras of $\map(M^d;\gl_N)$
($\gl_N$ the algebra of complex $N\times N$--matrices) for some $N$, and
elements $X$ in $\map(M^d;\gl_N)$ can be naturally identified with a
bounded operator on the Hilbert space
$\cH= L^2(M^d)\otimes V_{\rm spin}\otimes\C^N$
where $V_{\rm spin}$ is a vector space carrying the spin structure.
Physically $\cH$ is the one--particle Hilbert space of fermions on $M^d$,
and the Dirac operator $\D_A$ in an external Yang--Mills field $A$ is
a self--adjoint operator on $\cH$ playing the role of a
one--particle Hamiltonian.

The connection of this with NCG is as follows:
$\map(M^d;\gl_N)$ can be naturally embedded in the algebra $g_p$ of bounded
operators $u$ on $\cH$ satisfying the Schatten ideal condition that
$
\left(\ccr{\eps}{u}\ccr{\eps}{u}^*\right)^p
\mbox{ is trace class}
$
for $2p=d+1$ where $\eps=\sign(\D_{\, 0})$ is the sign of the free Dirac
operator $\D_{\, 0}$ on $M^d$ (see e.g.\ \cite{MR}), and these very algebras
$g_p$ play a fundamental role in NCG.  Moreover, the Yang--Mills field
configurations $A$ on $M^d$ can be embedded in the Grassmannian
$Gr_p$ of grading operators $F$ on $\cH$, $F=F^*=F^{-1}$, also satisfying
this Schatten ideal condition, and this embedding is given by
$A\mapsto F_A = \sign(\D_A)$ \cite{MR}.  Physically the Schatten ideal
condition can be regarded as a characterization of the degree of
ultra--violet (UV) divergence arising in the fermion sector of
$(d+1)$--dimensional Yang--Mills--fermion systems, and $2p=d+1$ corresponds
to the fact that UV divergences are worse in higher dimensions.

It is then natural to develop the representation theory for $g_p$ as a
whole and obtain the ones for $\map(M^d;G)$ by restriction from that as it
is the Schatten ideal condition which determines the appropriate
regularization procedure required to construct the operators representing
$\map(M^d;\gl_N)$.  Also the algebra $g_p$ contains other operators of
interest to quantum field theory, thus considering $g_p$ as a whole is not
only natural mathematically but also very useful from a physical point of
view.

This is well--established for $g_1$ which can be represented as current
algebra in a fermion Fock space with fermion currents constructed by normal
ordering (see e.g.\ \cite{PS,CR,M1,GL}).  From this one obtains as special
cases the wedge representations of affine Kac--Moody algebras (central
extensions of $\map(S^1;g)$) and also the Virasoro algebra which
played a central role in recent developments in (1+1) dimensional quantum
field theory.  For $g_{p\geq 2}$ analog representations in a fermion Fock
space have been found \cite{MR,M2}.  In this case, normal ordering of the
fermion currents is not sufficient but an additional multiplicative
regularization is required \cite{M3,L}.

The regularizations necessary to construct the fermion currents
representing $g_p$ lead to non--trivial 2--cocyles of the Lie algebras
$g_p$ in the current--current commutator relations.  Physically these
cocycles are Schwinger terms, and it should be possible to trace back all
anomalies of fermion--Yang--Mills systems
\cite{Jackiw} to these cocycles.  At least this is known to be the case
for (1+1) dimensions corresponding to $g_1$ (see e.g.\ a recent
construction of QCD$_{1+1}$ with massless quarks exploiting the
representation theory of $g_1$ and deriving all anomalies from the
$g_1$--cocycle \cite{LS}), and recently it has been shown by explicit
calculation that the cocycle of $g_3$ leads to the Gauss law anomaly
of (3+1)--dimensional chiral QCD \cite{LM}.

The technical difficulty in proving that the 2--cocycle of $g_p$ is indeed
equivalent to an anomaly in Yang--Mills gauge theory is that the former is
given by a Hilbert space trace of operators --- which in general is a
highly non--local expression --- whereas the latter are local, i.e.\
integrals of de Rham forms on some manifold $M^d$.  For example, the
2--cocycle of $g_1$ (originally found by Lundberg \cite{Lund}) is
$\hat{c}_1(u,v) = \f{1}{4}\Tra{\eps\ccr{\eps}{u}\ccr{\eps}{v}}$ where
$u,v\in\g_1$.  As will become clear below, it is actually more natural to
write this cocycle as
\alpheqn
\eq
\label{hc1}
\hat{c}_1(u,v) = \f{1}{2}\TraC{u\ccr{\eps}{v}}.
\eqend
where ${\rm Tr}_C$ is the {\em conditional trace} defined by
$\TraC{a}\equiv \f{1}{2}\Tra{a + \eps a\eps}$.
The corresponding 2--cocycle of $\map(M^1;g)$ is
\eq
\label{c1}
c_1(X,Y) = \f{1}{2\pi}\int_{M^1} \trac{X\dd Y}
\eqend
where ${\rm tr}_N$ is the usual trace of $N\times N$--matrices.  At first
sight it seems difficult to relate $\hat{c}(X,Y)$ to $c(X,Y)$.
Nevertheless, comparing \Ref{hc1} with \Ref{c1} is very suggestive: in NCG
the commutator $\ii\ccr{\eps}{\cdot}$ is the generalization of the exterior
differentiation $\dd(\cdot)$ of de Rham forms \cite{Connes}, thus \Ref{hc1}
looks exactly like the non--commutative generalization of \Ref{c1} if one
regards $\TraC{\cdot}$ as the non--commutative generalization of
integration $\f{1}{\pi}\int_{M^1}\tra{\cdot}$ of $\gl_N$--valued de Rham
forms on $M^1$.  Indeed, it is known that (see e.g.\ \cite{CR})
\eq
\label{int1}
\ii\TraC{X_0\ccr{\eps}{X_1}} = \f{1}{\pi} \int_{\R}\tra{X_0\dd X_1}
\quad \forall X_0,X_1\in\Map(\R;\gl_N)
\eqend
(and similarly for $M^1=S^1$) which completely justifies this point of view
and proves that $\hat{c}(X,Y)=c(X,Y)$ for $X,Y\in\Map(\R^1;g)$.
\reseteqn

In (3+1)--dimensions the generalization of \Ref{c1} is \cite{MR}
\alpheqn
\eq
\label{c3}
\hat c_3(u,v;F) =
-\f{1}{8}\TraC{(F-\eps)\ccr{\ccr{\eps}{u}}{\ccr{\eps}{v}} }
\eqend
where $u,v\in g_3$ and $F\in Gr_3$, and for $u=X$, $v=Y$
($X,Y\in\map(M^3;g)$) and $F=\sign(D_A)=F_A$ it is cohomologous to
the Gauss law anomaly \cite{M,FS}
\eq
\label{hc3}
c_3(X,Y;A) =\f{\ii}{24\pi^2} \int_{M^3} \trac{A\ccr{\dd X}{\dd Y}}
\eqend
which was shown in \cite{LM}.  Again seems natural to regard \Ref{hc3} as
non--commutative generalization of \Ref{c3} if one interprets $F-\eps$ as
the generalization of the 1--form $A$ (for a more detailed discussion see
\cite{LM}).  Especially, for $A$ a pure gauge we have $A= -\ii U^{-1}\dd U$
for some $U\in\map(M^3;G)$, hence $F_A = U^{-1}\ccr{\eps}{U}$, and \Ref{hc3}
would be indeed the non--commutative generalization of \Ref{c3} if we could
regard $\TraC{\cdot}$ as extension of
$\f{\ii}{3\pi^2}\int_{M^3}\tra{\cdot}$, i.e.\ (for $M^3=\R^3$)
\eqa
\label{int3}
\ii^3\TraC{X_0\ccr{\eps}{X_1}\ccr{\eps}{X_2}\ccr{\eps}{X_3}}
= \f{\ii}{3\pi^2}\int_{\R^3}\trac{X_0\dd X_1\dd X_2\dd
X_3}
\nonu
\quad \forall X_0,X_1,X_2,X_3\in\Map(\R^3;\gl(N)).
\eqaend
\reseteqn

In this paper we give a simple proof of the eqs.\ \Ref{int1} and
\Ref{int3} and their generalizations to arbitrary (even and odd) dimensions
$d$.
(In even dimensions $d$, the non--commutative integration is $\propto
\TraC{\gamma_{d+1}\cdot}$ involving a spin matrix $\gamma_{d+1}$
($\gamma_5$ for $d=4$) anticommuting with $\D_{\, 0}$.)

The method of proof is very simple and was inspired by the calculation in
\cite{LM}: we introduce a regularized
trace ${\rm Tr}_\Lam$ with a `momentum cut--off' $\Lam$ such that
$\TraL{a}$ exists and converges to $\TraC{a}$ as $\Lam\to\infty$ for
all conditionally trace--class operators $a$ on $\cH$.  As was shown in
\cite{LM}, one can easily calculate expressions of the form
$\TraL{\ccr{a}{b}}$ in the limit $\Lam\to\infty$ using symbol calculus of
PDOs \cite{symbol} (see also \cite{f}).  If $a$ is trace--class this is
obviously zero, but it is (in general) non--zero if $ab$ is only
conditionally trace--class.  In this case it is essentially a `surface
integral in Fourier space' which involves only the operators at large
momenta (= Fourier variables) and therefore can be calculated using
asymptotic expansions of the operator $\ccr{a}{b}$ in inverse powers of the
momenta.

It is worth noting that these `surface integral in Fourier space' (see \
eqs.\ \Ref{surface} and \Ref{equ}) have exactly the form typical for
Feynman diagrams giving anomalies (see e.g.\ \cite{JJ}).  Moreover, these
expressions have also a deep mathematical meaning: as shown in \cite{f},
for PDOs $a,b$ such that $\ccr{a}{b}$ is conditionally trace class,
$\lim_{\Lam\to\infty} \TraL{\ccr{a}{b}}$ is equal (up to a constant) to the
Wodzicki residue ${\rm Res}\left(\ccr{\log(|\D_{\, 0}|)}{a}{b}\right)$ \cite{W}
playing an important role in NCG.

We now write (for $d=3$)
\[
\TraL{X_0\ccr{\eps}{X_1}\ccr{\eps}{X_2}\ccr{\eps}{X_3}} =
\TraL{\ccr{X_0\ccr{\eps}{X_1}\ccr{\eps}{X_2}\eps}{X_3}}
- \TraL{\ccr{X_0\ccr{\eps}{X_1}\ccr{\eps}{X_2}}{X_3}\eps}
\]
and with the arguments given in \cite{LM} (using the calculus of PDOs) it
is not difficult to show that the first term on the r.h.s.  in the limit
$\Lam\to\infty$ is equal to the r.h.s.  of \Ref{int3} (we will, however,
give more detailed argument for this in the present paper).  The difficult
part in the proof of \Ref{int3} is to show that second term on the r.h.s.
of this eq.\ actually is zero.

The result of this paper provides a natural generalization of integration
of de Rham forms to NCG.  It is valid for a differential complex over
$g_p$ with differentiation given by $\ii\ccr{\eps}{\cdot}$.  We note that
another such generalization was given in \cite{Connes1}.  In this case, the
non--commutative differentiation is defined by $\ii\ccr{\D_{\,
0}}{\cdot}$, and the integration is in terms of the Dixmier trace rather
than the ordinary Hilbert space trace as in our case.  We note that for the
differential complex with differentiation $\ii\ccr{\eps}{\cdot}$, the
algebra product is equal to the product of Hilbert space operators (this
follows from $\eps^2=1$ --- see Section 4).  It therefore seem more natural
than the one with differentiation $\ii\ccr{\D_{\, 0}}{\cdot}$ which does
not have this property.

Our proof is restricted to manifolds $M^d=\R^d$.  It is natural to
conjecture a similar result for arbitrary Riemannian manifolds $M^d$
allowing for a spin structure, but we have not been able to prove this in
general.  We note that Connes' non--commutative integration \cite{Connes1}
generalizes integration of de Rham forms for all such manifolds $M^d$.

The plan of this paper is as follows.  We introduce the notation and state
the result in the next section.  The proof is then given in Section 3.  It
is divided in several Lemmas which are proved in very detail using only
elementary arguments.  We believe that this is justified because these
Lemmas also play a fundamental role in other applications of NCG to quantum
field theory and these arguments should therefore be useful, at least for
physicists.  It is also intended to make explicit that the crucial steps in
the proof are very similar to arguments used standard perturbative
calculations of anomalies in particle physics.  We end with an outline how
our result fits into the general framework of NCG and comments on other
possible applications to quantum field theory in Section 4.

\section{Notation and Result}
For $\cH$ a separable Hilbert space, we denote as $\BO$ and $\TR$ the
bounded and trace--class operators on $\cH$, respectively, and $\SIq=\left\{
a\in\BO\left| (a^*a)^{q/2}\in\TR\right.  \right\}$ are the Schatten classes
($q\in\N$; $*$ is the Hilbert space adjoint).  We note that for
$a_0,a_1,\ldots a_d\in\SIq$, $a_0a_1\cdots a_d$ is trace class for $d+1\geq
q$ \cite{Simon}.

Let
\alpheqn
\eq
\cH=L^2(\R^d)\otimes\C^\nu\otimes\C^N
\eqend
with
\eq
\nu=2^{[d/2]}
\eqend
\reseteqn
where $[d/2]=(d-1)/2$ for $d$ odd and $[d/2]=d/2$ for $d$ even. The free
Dirac operator $\D_{\, 0}$ on $\R^d$ is
\alpheqn
\eq
\label{16a}
\D_{\, 0}=\sum_{i=1}^d \gamd_i(-\ii)\f{\partial}{\partial x_i}
\eqend
where $\gamd_i$ are self--adjoint
$\nu\times\nu$ matrices acting on $\C^\nu$ and obeying
\eq
\label{16b}
\gamd_i\gamd_j +\gamd_j\gamd_i = 2\delta_{ij}1_{\nu\times\nu}
\eqend
\reseteqn
with $1_{\nu\times\nu}$ the $\nu\times\nu$ unit matrix.
This naturally defines a self--adjoint operator on $\cH$ which we denote
by the same symbol $\D_{\, 0}$.

To be specific we choose the following representation for the
$\gamma$--matrices: for $d=1$ we have $\nu=1$ and
\alpheqn
\eq
\gamma_1^{(1)}\equiv 1,
\eqend
and for all \ul{odd $d$} we define inductively ($0_{\nu\times\nu}$
is the $\nu\times\nu$ matrix with all matrix elements $=0$)
\eqa
\label{17a}
\gazd_i\equiv\left(\bma{cc} 0_{\nu\times\nu}& \gamd_i\\ \gamd_i&
0_{\nu\times\nu} \ema\right) \quad\mbox{ for $i=1,2,\ldots d$,}\nonu
\gazd_{d+1}\equiv\left(\bma{cc} 0_{\nu\times\nu}& -\ii 1_{\nu\times\nu}
\\ \ii 1_{\nu\times\nu} & 0_{\nu\times\nu} \ema\right), \qquad
\gazd_{d+2}\equiv\left(\bma{cc} 1_{\nu\times\nu}& 0_{\nu\times\nu}\\
0_{\nu\times\nu}& -1_{\nu\times\nu} \ema\right).
\eqaend
For all \ul{even $d$} we choose
\eq
\label{17b}
\gamd_i\equiv \gamma^{(d+1)}_i\quad\mbox{ for $i=1,2,\ldots d+1$},
\eqend
\reseteqn
hence there is an additional spin matrix $\gamd_{d+1}$
which can be identified with a grading operator in
$\cH$ (by abuse of notation, we use the same symbol for
$\gamd_{d+1},1_{\nu\times\nu}\in\gl_N$ and the corresponding
operators on $\cH$).

Then there is a natural embedding\footnote{to avoid confusion, we
distinguish here $X\in\Map(\R^d;g)$ from the corresponding operator $\hat
X$ on $\cH$} $X\to\hat X$ of $\Map(\R^d;\gl_N)$
in $\BO$,
\eq
\label{embed0}
(\hat X f)(x) \equiv X(x) f(x)
\quad \forall f\in\cH
\eqend
(we write $\cH\ni f:\R^d\to \C^\nu\otimes\C^N,
f\mapsto f(x)$, and $\gl_N$ acts naturally on $\C^N$).

Using the spectral theorem for self--adjoint operators \cite{RS1} we define
$\eps = \sign(\D_{\, 0})$ where $\sign(x) = +1 (-1)$ for $x\geq 0$ ($x<0$).

The Schatten ideal discussed in the introduction can be written as
\eq
\label{embed}
X\in\Map(\R^d;\gl_N) \Rightarrow \ccr{\eps}{\hat X}\in\SIq \mbox{ if }
q = d+1
\eqend
(see e.g.\ \cite{MR}).

With ${\rm Tr}$ the usual Hilbert space trace, we note that
the {\em conditional trace}
\eq
\label{16}
\TraC{a} \equiv\half\Tra{a+\eps a\eps}
\eqend
is well-defined for all $a$ in the {\em conditional trace class}
\eq
\TRC \equiv \left\{ a\in\BO| a+\eps a\eps\in\TR\right\} .
\eqend

We can now state our result.

{\bf Theorem:}
Let
\eq
\Gamma^{(d)}\equiv\left\{\bma{ll} 1_{\nu\times\nu}
&\mbox{ if $d$ is odd}\\ \gamd_{d+1}
& \mbox{ if $d$ is even}\ema\right. .
\eqend
Then for all $X_0,X_1,\ldots X_d\in\Map(\R^d;\gl_N)$,
\eq
\label{bb}
\Gamma^{(d)} \hat X_0\ccr{\eps}{\hat X_1} \cdots \ccr{\eps}{\hat X_d} \in\TRC,
\eqend
and
\alpheqn
\eq
\label{result}
\ii^d\TraC{\Gamma^{(d)}
\hat X_0\ccr{\eps}{\hat X_1}\cdots \ccr{\eps}{\hat X_d}} =
c_d \intS\tra{X_0 \dd X_1\cdots \dd X_d}
\eqend
with a normalization constant
\eq
\label{cd}
c_d = (2\ii)^{[d/2]} \f{1}{d(2\pi)^d}
\f{2\pi^{d/2}}{\Gamma(d/2)} .
\eqend
\reseteqn

For example
\eq
c_1 = \f{1}{\pi}, \quad
c_2 = \f{\ii}{2\pi}, \quad
c_3 = \f{\ii}{3\pi^2}, \quad
c_4 = \f{-1}{2\pi^2}\quad \ldots
\eqend

\section{Proof}

To see that \Ref{bb} is true we observe that
--- denoting the l.h.s. of \Ref{bb} as $a$ ---
\[
a +\eps a\eps = \Gamma^{(d)}\eps\ccr{\eps}{\hat X_0} \ccr{\eps}{\hat X_1}
\cdots \ccr{\eps}{\hat X_d}
\]
(to see this use repeatedly
$\eps\ccr{\eps}{\hat X_i}\eps = -\ccr{\eps}{\hat X_i}$ and $\eps^2=1$;
for even $d$ one also needs
$\eps \Gamma^{(d)}= - \Gamma^{(d)}\eps$) which is trace--class
by \Ref{embed}.

For $\Lam>0$ and $a\in\TRC$ we define the cutoff trace
\eq
\TraL{a} = \Tra{a P_\Lam},\quad P_\Lam = \Theta(\Lam-|\D_{\, 0}|)
\eqend
where $\Theta(\Lam-|x|) = 1$ for $|x|\leq \Lam$ and 0 otherwise.

{\bf Lemma 1:} For all $a\in\TRC$, $aP_\Lam$ is trace class
for all $\Lam<\infty$, and
\eq
\TraC{a} = \lim_{\Lam\to\infty} \TraL{a} .
\eqend

{\em Proof:} If $a\in\TR$ then trivially $a P_\Lam\in\TR$ and $\Tra{a} =
\lim_{\Lam\to\infty}\TraL{a}$ as $P_\Lam$ converges strongly to the
identity for $\Lam\to\infty$.  If $a\in\TRC$ then $\f{1}{2}\TraL{a + \eps
a\eps}=\TraL{a}$ as $\eps$ commutes with $P_\Lam$.  By definition of ${\rm
Tr}_C$ and $\TRC$ this implies the Lemma.

{\bf Lemma 2:} Let $\tras{\cdot}$ be the trace of $\nu\times\nu$ matrices
acting on $\C^\nu$ and $i_1,i_2,\ldots i_{d-2n}\in\{1,2,\ldots
d\}$. Then for $n=0,1,\ldots [d/2]$,
\eq
\label{A1}
\tras{\Gamma^{(d)} \gamd_{i_1}\gamd_{i_2}\cdots\gamd_{i_{d-2n} }} =
\delta_{n,0}(2\ii)^{[d/2]}
\epsilon_{i_1 i_2\cdots i_n}
\eqend
where $\epsilon_{i_1 i_2\cdots i_n}$ is the anti-symmetric tensor (i.e.\
it is equal to $+1$ ($-1$) for $(i_1,i_2,\ldots i_d)$ an even (odd)
permutation of $(1,2,\ldots d)$ and $0$ otherwise).

{\em Proof:} If $d=2m$ is even then the eqs.\ \Ref{A1} are just special cases
of
the eqs.\ for $d=2m+1$, hence we can restrict ourselves to odd $d$'s. For
$d=1$ eq.\ \Ref{A1} is true trivially. For general \ul{odd} $d$ we prove it by
induction. We consider
\[ M= \gazd_{i_1}\gazd_{i_2}\cdots\gazd_{i_{d+2-2n} }\]
for $n=0,1,\ldots (d+1)/2$.  Without loss of generality we first assume that
$1\leq i_1<i_2<\ldots <i_{d+2-2n}\leq d+2$.  (This is because, by using the
relation \Ref{16b}, $M$ can be always brought to the form
$\pm\gazd_{j_1}\gazd_{j_2}\cdots\gazd_{j_{d+2-2n}}$ with $1\leq j_1\leq
j_2<\ldots \leq j_{d+2-2n}\leq d+2$.  If two or more of these indices are
equal, $\left(\gazd_j\right)^2=1_{2\nu\times 2 \nu}$ implies that $M$
is equal to some matrix $\pm \gazd_{k_1} \gazd_{k2}
\cdots\gazd_{k_{d-2m}}$
with $m>n$ and $1\leq k_1< k_2<\ldots <k_{d+2-2m}\leq d+2$.)

\newcommand{\traz}[1]{{\rm tr}_{2\nu}\left({#1}\right)}

If $i_{d+2-2n}=d+2$, $i_{d+1-2n}=d+1$ then eq.\ \Ref{17a} implies
\[
M = \left(\bma{cc} \ii \gamd_{i_1}\cdots\gamd_{i_{d-2n}}&
0_{\nu\times\nu} \\ 0_{\nu\times\nu}&\ii \gamd_{i_1}\cdots\gamd_{i_{d-2n}}
\ema \right),
\]
hence $\traz{M} = (2\ii)\tras{\gamd_{i_1}\cdots\gamd_{i_{d-2n}}}$,
and for $n\geq 1$ $\traz{M} = 0$ follows from the induction hypothesis.
If $i_{d+2-2n}\neq d+2$, or $i_{d-2n+2}=d+2$ and $i_{d-2n+1}\neq d+1$, then
$\traz{M} = 0$ trivially because $M$ is of the form
\[
\left(\bma{cc} 0_{\nu\times\nu}& (\cdots) \\ \pm(\cdots)&
0_{\nu\times \nu}\ema \right) \quad \mbox{ or }\quad
\left(\bma{cc} (\cdots) &0_{\nu \times \nu} \\
0_{\nu \times \nu} & -(\cdots) \ema\right).
\]
This proves \Ref{A1} for $n\geq 1$. To prove it for $n=0$ we observe that
by the relations above and induction,
\nonueqa
\traz{\gazd_1\cdots\gazd_{d+2} } = (2\ii)\tras{\gamd_1\cdots\gamd_d} \\
= \cdots = (2\ii)^{(d+1)/2}{\rm tr}_1\left(\gamma^{(1)}_1\right) =
(2\ii)^{(d+1)/2}.
\nonueqaend
Moreover (as discussed above --- we now allow for arbitrary indices
$i_1,i_2\cdots i_{d+2}$),
\[
\tras{\gazd_{i_1}\gazd_{i_2}\cdots\gazd_{i_{d+2}}}
\]
is non-zero only if $(i_1,i_2,\ldots i_{d+2})$ is a permutation of
$(1,2,\ldots d+2)$, and if this is the case it is equal to
$(+/-)$ $\tras{\gazd_{1}\gazd_{2}\cdots\gazd_{d+2}}$
depending on whether this permutation is even$/$odd. This implies
\Ref{A1} for $n=0$ and completes the proof of Lemma 2.

{\bf Lemma 3:} Let $X_1,\ldots X_n\in \Map(\R^d;\gl_N)$. Then for
all $n=0,1,\ldots [d/2]$ and $\Lam>0$,
\eq
\label{A2}
\TraL{\Gamma^{(d)} \hat X_1\eps \hat X_2\eps \cdots \hat X_{d-2n} \eps} =
0 .
\eqend

{\em Proof:} We first introduce some notation. For $L^2$--functions $f$ on
$\R^d$ we define the Fourier transform as
\[
\tilde f(p) = \intS \dvx\ee{\ii px} f(x)
\]
where $px=\sum_{i=1}^d p_i x_i$ and $\dvx= d^d x$. The integral
kernel $K(a)$ of $a\in\BO$ is the $\gl_\nu\otimes\gl_N$--valued function
on $\R^d\times\R^d$ such that
\eq
\label{Ka}
\wt{(af)}(p)=\intS\dvq K(a)(p,q)\tilde f(q)\quad \forall f\in\cH
\eqend
where $\dvq=d^d q/(2\pi)^d$. Then for all $a,b\in\BO$,
\[
K(ab)(p,q) = \intS\dvk K(a)(p,k)K(b)(k,q),
\]
and for all $a\in\TR$,
\[
\Tra{a} = \intS\dvp\, \tra{K(a)(p,p)}
\]
where ${\rm tr}={\rm tr}_\nu {\rm tr}_N$.

Therefore, as
\nonueqa
K(\eps)(p,q) &=& (2\pi)^d\delta^d(p-q)\eps(p)\\
K(\hat X)(p,q) &=& \tilde X(p-q)\\
K(P_\Lam)(p,q) &=& (2\pi)^d\delta^d(p-q)\Theta(\Lam -|p|)
\nonueqaend
with
\eq
\label{eps}
\eps(p) \equiv \f{1}{|p|} p\slD \equiv \f{1}{|p|}\sum_{i=1}^d p_i\gamd_i,
\eqend
we get
\nonueqa
\TraL{\Gamma^{(d)} \hat X_1\eps \hat X_2\eps \cdots \hat X_n\eps} =
\intF\dvp\intS\dvq_1\cdots\intS\dvq_n
\\ \times
\tra{\Gamma^{(d)}\hat
X(p-q_1)\eps(q_1)\hat X_2(q_1 -q_2)\eps(q_2)\cdots \hat
X_n(q_{n-1}-p)\eps(p)}
\nonueqaend
with $B_\Lam^d= \left\{\left. p\in\R^d\right| |p|\leq\Lam \right\}$,
and \Ref{A2} trivially follows from \Ref{A1} for $n<d$.

The only non--trivial case is $n=d$. Shifting the integration variables to
\[
Q_1=p-q_1,\quad Q_2=q_1-q_2,\quad \ldots \quad Q_{d-1}=q_{d-2} -q_{d-1}
\]
we can write
\nonueqa
\TraL{\Gamma^{(d)} \hat X_1\eps \hat X_2\eps \cdots \hat X_d\eps} =
\intS\dvQ_1\cdots\intS\dvQ_{d-1}\\ \times \trac{\hat X_1(Q_1)\hat X_2(Q_2)
\cdots
\hat X_{d-1}(Q_{d-1})\hat X_d(-Q_1-Q_2-\ldots -
Q_{d-1})} f_\Lam(Q_1,Q_2,\ldots Q_{d-1})
\nonueqaend
where
\[
f_\Lam(Q_1,Q_2,\ldots Q_{d-1}) =
\intF\dvp\;\tras{\Gamma^{(d)}\eps(p-Q_1)\eps(p-Q_1-Q_2)\cdots\eps(p-Q_1-\ldots
-Q_{d-1})\eps(p)}.
\]
To evaluate this latter function we use the representation
\[
\eps(Q) =\int_{\R}\f{dt}{\sqrt{2\pi}}\ee{-\half t^2 Q^2} \sQ
\]
($Q^2=\sum_{i=1}^{d}Q_i^2$), thus
\nonueqa
f_\Lam(Q_1,Q_2,\ldots Q_{d-1}) =
\int_{\R}\f{dt_1}{\sqrt{2\pi}}\ee{-\half t_1^2 Q_1^2}
\int_{\R}\f{dt_2}{\sqrt{2\pi}}\ee{-\half t_2^2 (Q_1+Q_2)^2}\ldots \\ \times
\int_{\R}\f{dt_{d-1}}{\sqrt{2\pi}}\ee{-\half t_{d-1}^2 (Q_1+\ldots
+Q_{d-1})^2}\int_{\R}\f{dt_d}{\sqrt{2\pi}}(\cdots)
\nonueqaend
where
\[
(\cdots) = \intF\dvp\ee{-\half t^2 p^2}\ee{pv}\, \tras{\Gamma^{(d)}
(p\slD -\sQ_1)(p\slD -\sQ_1 -\sQ_2)\cdots (p\slD -\sQ_1 -\ldots
-\sQ_{d-1}) p\slD}
\]
with $t^2=\sum_{i=1}^d t_i^2$, $pv=\sum_{i=1}^dp_i v_i$, and
\[
v= t_i^2 Q_1 + t_2^2(Q_1+Q_2) + \cdots + t_{d-1}^2(Q_1+Q_2+ \ldots +
Q_{d-1}) .
\]
Using \Ref{A1} we see that $(\cdots)$ is proportional to
\[
\intF\dvp\ee{-\half t^2 p^2}\ee{pv}\sum_{i_1 i_2\ldots i_d}
\epsilon_{i_1 i_2\ldots i_d}
(p-Q_1)_{i_1}(p-Q_1-Q_2)_{i_2}\cdots (p-Q_1-Q_2-\ldots
-Q_{d-1})_{i_{d-1}}p_{i_d}
\]
which due to the antisymmetry of the $\epsilon$-symbol is identical with
\[
(-)^{d-1} \sum_{i_1 i_2\ldots i_d}
\epsilon_{i_1 i_2\ldots i_d}(Q_1)_{i_1}(Q_2)_{i_2}\cdots
(Q_{d-1})_{i_{d-1}} \intF\dvp\ee{-\half t^2 p^2}\ee{pv}p_{i_d}.
\]
Obviously the integral $\intF\dvp\ee{-\half t^2 p^2}\ee{pv}p$ is a vector
in $\R^d$ parallel to $v$, and as $v$ is just a linear combination of the
$Q_\nu$ for $\nu=1,2,\ldots (d-1)$ the anti-symmetry of the
$\epsilon$-symbols implies $(\cdots)=0$ and thus \Ref{A2} for $n=d$.

{\bf Lemma 4:}  For all $X_0,X_1,\ldots X_d\in\Map(\R^d;\gl_N)$,
\eq
\label{A3}
\TraL{\Gamma^{(d)}\hat X_0\ccr{\eps}{\hat X_1}\cdots\ccr{\eps}{\hat X_d}} =
\TraL{\ccr{\Gamma^{(d)}\hat X_0\ccr{\eps}{\hat X_1}
\cdots\ccr{\eps}{\hat X_{d-1}}\eps}{\hat X_d} }.
\eqend

{\em Proof:} The difference of the r.h.s.  and the
l.h.s.  of \Ref{A3} is equal to
\[
\TraL{\ccr{\Gamma^{(d)}\hat X_0\ccr{\eps}{\hat X_1}
\cdots\ccr{\eps}{\hat X_{d-1}}}{\hat X_d}\eps}
\]
which is just a linear combination of terms $\TraL{\Gamma^{(d)} \hat
Y_1\eps \hat Y_2\eps \cdots \hat Y_{d-2n} \eps}$ with $n\leq d$ and
$Y_i\in\Map(\R^d;\gl_N)$ (use $\eps^2=1$; for even $d$ one also needs
$\Gamma^{(d)}\eps = -\eps\Gamma^{(d)}$ and cyclicity of ${\rm tr}_\nu$) and
therefore zero by Lemma 3.

{\bf Proof of the Theorem:} The Lemmas above imply,
\eq
\label{a}
\TraC{\Gamma^{(d)}\hat X_0\ccr{\eps}{\hat X_1}\cdots\ccr{\eps}{\hat X_n}} =
\llim \TraL{\ccr{\Gamma^{(d)}\hat X_0\ccr{\eps}{\hat X_1} \cdots\ccr{\eps}{\hat
X_{d-1}}\eps}{X_d}}.
\eqend
We now evaluate the r.h.s. of this
using symbol calculus \cite{symbol}. We recall that every pseudo
differential operator (PDO\footnote{all operators of interest to us are
PDOs}) $a$ on $\cH$ can be represented by its {\em symbol} $\sig(a)(p,x)$
which is a $\gl_\nu \otimes\gl_N$-valued function on $\R^d\times\R^d$ and
defined
such that for any $f\in \cH$,
\eq
\label{symb}
(af)(x) = \int\dvp
\ee{-\ii px}\sig(a)(p,x)\tilde f(p)
\eqend
where $\tilde f(p)$
is the Fourier transform of $f$. It follows then that
\eq
\label{product}
\sig(ab)(p,x)= \intS\dvq\intS\dvy\ee{\ii(x-y)(p-q)}
\sig(a)(q,x)\sig(b)(p,y),
\eqend
and for $a$ trace--class,
\[
\Tra{a} =
\intS\dvp\intS\dvx\, \tra{\sig(a)(p,x)}.
\]
Especially, $\sig(\eps)(p,x)=\eps(p)$ and $\sig(\hat X)(p,x)=X(x)$ for all
$X\in\Map(\R^d;g)$.  Moreover, $\sig(P_\Lam)(p,x) = \Theta(\Lam - |p|)$,
hence
\eq
\TraL{a} =
\intF\dvp\intS\dvx\, \tra{\sig(a)(p,x)}.
\eqend

All operators $a$ of interest to us allow an asymptotic expansion
$\sig(a)\sim\sum_{j=0}^{\infty}\sig_{-j}(a)$ where $\sig_{-j}(a)(p,x)$ is
homogeneous of degree $-j$ in
$p$ (i.e.\ $\sig_{-j}(a)(sp,x)=s^{-j}\sig_{-j}(a)(p,x)$ for
all $s>0$) and goes to zero like $|p|^{-j}$ for $|p|\to\infty$.
We write
\eq
\sig(a)(p,x)=\sum_{j=0}^{n}\sig_{-j}(a)(p,x) + \OO(|p|^{-n-1}).
\eqend
Moreover, eq.\ \Ref{product} has an asymptotic
expansion in powers of $|p|^{-1}$,
\eq
\label{as}
\sig(ab)(p,x) \sim \sum_{n=0}^{\infty}\sum_{i_1\ldots i_n = 1}^d
\f{(-\ii)^n}{n!}
\f{\partial^n\sig(a)(p,x)}{\partial p_{i_1}\cdots\partial p_{i_n}}
\f{\partial^n\sig(b)(p,x)}{\partial x_{i_1}\cdots\partial x_{i_n}}.
\eqend
This allows to determine the asymptotic expansion of $\sig(ab)$ from the ones
of $\sig(a)$ and $\sig(b)$.  Especially if $\sig(a)$ is $\OO(|p|^{-n})$ and
$\sig(b)$ is $\OO(|p|^{-m})$ then $\sig(ab)$ is $\OO(|p|^{-(n+m)})$.

For PDOs $a\in\TR$ and $X\in\Map(\R^d;\gl_N)$, $\TraL{\ccr{a}{\hat X}}$
converges to zero for $\Lam\to\infty$ (as $P_\Lam$ strongly converges to
the identity this trivially follows from the the
cyclicity of trace).  Especially this is true if $\sig(a)$ is
$\OO(|p|^{-d-1})$.  If $\sig(a)$ is only $\OO(|p|^{-d+1})$ then $a$ is not
trace class but $\TraL{\ccr{a}{\hat X}}$ still has a well--defined and (in
general) non--trivial limit. Indeed, eq.\ \Ref{as} suggests that
\eq
\label{aX}
\TraL{\ccr{a}{\hat X}} = (-\ii)\intF\dvp\intS\dvx\sum_{i=1}^d
\tra{\f{\partial\sig(a)(p,x)}{\partial p_i}\f{\partial X(x)}{\partial
x_i}} +\OO\left(\Lam^{-1}\right) .
\eqend
This is true as the terms neglected in the asymptotic expansion of the
operator products are also total derivatives and, by Stokes theorem
\Ref{stokes}, can be written as surface integrals over $|p|=\Lam$ which vanish
as
$\Lam\to\infty$  (the details of this argument are given in Appendix A).

Using Stokes' theorem
\eq
\label{stokes}
\intF\dvp\, \f{\partial}{\partial p_i}\, f(p) = \intS\dvp\,\delta(\Lam
-|p|)\f{p_i}{|p|} f(p)
\eqend
(which in the present case is just equivalent to
\[
\intS\dvp \f{\partial}{\partial p_i} \,\Theta(\Lam -|p|)f(p) = 0 )
\]
we therefore get
\eq
\label{equ}
\TraL{\ccr{a}{\hat X}} = (-\ii)\sum_{i=1}^d
\intS\dvp\f{p_i}{|p|}\delta(\Lam-|p|) \intS\dvx
\tra{\sig(a)_{-d+1}(p,x)\f{\partial X(x)}{\partial x_i}} +
\OO\left(\Lam^{-1}\right)
\eqend
where we replaced $\sig(a)$ by its $\OO(|p|^{-d+1})$--piece $\sig_{-d+1}(a)$
as only this gives a non--zero contribution for $\Lam\to\infty$.

For $X\in\Map(\R^d;\gl_N)$, eq.\ \Ref{as} implies
\[
\sig\left(\ccr{\eps}{\hat X}\right)(p,x) = (-\ii)\sum_{i=1}^d \f{\partial
\eps(p)}{\partial p_i}\f{\partial X(x)}{\partial x_i} + \OO(|p|^{-2}),
\]
hence $a=\Gamma^{(d)}\hat X_0\ccr{\eps}{\hat X_1}
\cdots\ccr{\eps}{\hat X_{n-1}}\eps$ is $\OO(|p|^{-d+1})$ and
\[
\sig(a)_{-d+1}(p,x) = (-\ii)^{d-1}\Gamma^{(d)}
X_0(x)\sum_{i_1\ldots i_{d-1}=1}^d \f{\partial
\eps(p)}{\partial p_{i_1}}\f{\partial X_1(x)}{\partial x_{i_1}}
\cdots \f{\partial
\eps(p)}{\partial p_{i_{d-1}}}\f{\partial X_{d-1}(x)}{\partial x_{i_{d-1}}}.
\]
Thus with \Ref{equ},
\nonueqa
\TraL{\ccr{\Gamma^{(d)}\hat X_0\ccr{\eps}{\hat X_1}
\cdots\ccr{\eps}{\hat X_{n-1}}\eps}{\hat X_n} }
=
\sum_{i i_1\ldots i_{d-1}=1}^d
(-\ii)^d I^\Lam_{i i_1\ldots i_{d-1}}
\\ \times
\intS\dvx \,\trac{X_0(x) \f{\partial
X_1(x)}{\partial x_{i_1}} \cdots \f{\partial X_{d-1}(x)}{\partial
x_{i_{d-1}}} \f{\partial X_{d}(x)}{\partial x_{i}} }
+\OO\left(\Lam^{-1}\right)
\nonueqaend
where
\[
I^\Lam_{i i_1\ldots i_{d-1}} = \intS\dvp\, \delta(\Lam-|p|)
\f{p_i}{|p|}\tras{\Gamma^{(d)}
\f{\partial \eps(p)}{\partial p_{i_1}} \cdots \f{\partial
\eps(p)}{\partial p_{i_{d-1}}}\eps(p)} .
\]
Now \Ref{eps} implies
$\partial\eps(p)/\partial p_i = \sum_{j=1}^d P_{ij}\gamd_j /|p|$ with
$P_{ij} = (\delta_{ij} - p_i p_j/|p|^2)$, hence
with Lemma 2 we get
\nonueqa
\tras{\Gamma^{(d)}
\f{\partial \eps(p)}{\partial p_{i_1}} \cdots \f{\partial
\eps(p)}{\partial p_{i_{d-1}}}\eps(p)} = \sum_{j_1\ldots j_d = 1}^d
(2\ii)^{[d/2]}\epsilon_{j_1\ldots j_d}
\f{1}{|p|^d} P_{i_1 j_1}\cdots P_{i_{d-1}j_{d-1}}p_{j_d} \\=
(2\ii)^{[d/2]}\sum_{j=1}^d
\epsilon_{i_1\ldots i_{d-1} j} \f{p_{j}}{|p|^d}
\nonueqaend
where we used the antisymmetry of the $\epsilon$--symbol. Thus
\[
I^\Lam_{ii_1\ldots i_{d-1}} = (2\ii)^{[d/2]}\sum_{j=1}^d
\epsilon_{i_1\ldots i_{d-1} j} \, (\cdots)_{ij}
\]
with
\[
(\cdots)_{ij} = \intS\dvp\, \delta(\Lam -|p|) \f{p_ip_j}{|p|^{d}} =
\delta_{ij} \f{1}{d(2\pi)^d}\f{2\pi^{[d/2]}}{\Gamma(d/2)};
\]
we rescaled $\xi=p/\Lam$ and used $\intS d\xi \,\delta(1-|\xi|) =
2\pi^{d/2}/\Gamma(d/2)$ (volume of the unit sphere $S^{d-1}$). This implies
\[
I^\Lam_{ii_1\ldots i_{d-1}} = c_d \epsilon_{i_1\ldots i_{d-1} i}
\]
with $c_d$ \Ref{cd}.
Putting these equations together we end up with
\nonueqa
\TraL{\ccr{\Gamma^{(d)}\hat X_0\ccr{\eps}{\hat X_1}
\cdots\ccr{\eps}{\hat X_{d-1}}\eps}{\hat X_d} }\\
= (-\ii)^d c_d\intS\dvx \,\sum_{i_1\ldots i_d=1}^d\epsilon_{i_1\ldots i_d}
\trac{X_0(x) \f{\partial
X_1(x)}{\partial x_{i_1}} \cdots \f{\partial X_{d}(x)}{\partial
x_{i_{d}}}} +\OO\left(\Lam^{-1}\right),
\nonueqaend
thus eq.\ \Ref{a} implies \Ref{result} which completes our proof.

\section{Final Comments}

Let $\cA$ be an associative algebra over $\C$.
The basic object of NCG is the {\em differential complex} $(\Omega,\dd)$
over $\cA$, which is a $\N_0$--graded complex algebra,
\[\Omega =\bigoplus_{n=0}^\infty \Omega^{(n)}\]
where $\Omega^{(n)}$ are $\cA$--bimodules and $\Omega^{(0)}=\cA$.
Moreover, there is a linear operator
\[
\dd: \Omega^{(n)}\to \Omega^{(n+1)}
\]
satisfying $\dd^2 =0$ and $\dd(\omega\omega')=(\dd\omega)\omega' +(-)^n
\omega(\dd\omega')$ for all $\omega\in\Omega^{(n)}, \omega'\in\Omega$.
Here we restrict ourselves to the case where for all $n$, $\Omega^{(n)}$ is
equal to the linear span of forms $u_0\dd u_1 \cdots \dd u_n$ with
$u_i\in\cA$, hence $\Omega$ is determined by $\cA$ and $\dd$.

The most prominent example is the de Rham complex $(\Omega_d,\dd)$ of forms on
$\R^d$ which is the differential complex over $\cA_d=\Map(\R^d;\gl_N)$ with
$\dd$ the exterior differentiation of forms. In this case
$\Omega_d^{(n)}=\emptyset$ for $n>d$.

Another important example is the differential complex $(\hat\Omega_p,\ddd)$
over $$g_p=\left\{\left. u\in\BO\right|  \ccr{\eps}{u}\in B_{2p} \right\}$$
where $2p\in\N$, $\cH$ a separable Hilbert space, $\eps$ a grading operator
on $\cH$ ($\eps=\eps^*=\eps^{-1}$), and
\eq
\ddd u = \ii\ccr{\eps}{u}\quad \forall u\in g_p.
\eqend
Then $\ddd\hat \omega = \ii(\eps \hat \omega -(-)^n \hat \omega\eps)$ for
all $\hat\omega\in\hat\Omega_p^{(n)}$ showing that the algebra product in
$\hat\Omega_p$ is equal to the product as operators in $\BO$.  Moreover,
$\hat\omega\in\hat\Omega_p^{(n)}$ is conditionally trace class for $n\geq
2p$ (as before, $\TraC{a}=\f{1}{2}\Tra{a+\eps a \eps}$).

The Schatten ideal condition shows that $\hat X\in\g_p$ for all
$X\in\cA_d$ if $2p=d+1$, hence $\Omega_d$
can be naturally embedded in $\hat\Omega_p$, i.e.\ the differential
complex $(\hat\Omega_p',\ddd)$ over $\cA_p'=\left\{\left. \hat X\right|
X\in\cA_d \right\}$ is a subcomplex of $(\hat\Omega,\ddd)$ for $2p=d+1$
and the linear map $c:\hat\Omega_p'\to\Omega_d$,
\eq
\label{c}
\hat \omega= \ii^d\hat X_0\ccr{\eps}{\hat X_1}\cdots \ccr{\eps}{\hat
X_n} \mapsto c(\hat\omega) = X_0\dd X_1\cdots \dd X_n
\eqend
is a homomorphism of differential complexes.
Thus $(\hat\Omega_p,\ddd)$ are natural generalizations of de Rham
complexes to NCG.

The de Rham complex has another important property, namely it allows for
an integration $\int$ which is the linear map $\Omega_d\to\C$ defined by
\eq
\int\omega = \left\{ \bma{cc} \int_{\R^d}\tra{\omega} & \mbox{ for
$\omega\in\Omega_d^{(d)}$} \\ 0 & \mbox{ otherwise} \ema \right. ,
\eqend
and Stokes theorem holds,  $\int\dd \omega = 0$ for all
$\omega\in\Omega_d$.

The theorem of the present paper shows that the natural generalization of
this integration of de Rham forms to NCG is the linear mapping
$\hat{\int}: \hat\Omega_p\to \C$, defined by
\eq
\hat{\int} \hat\omega = \left\{ \bma{cc} c_{2p}^{-1}
\TraC{\Gamma^{2p}\hat\omega} &\mbox{ for
$\hat\omega\in \hat\Omega^{n}_p$, $n\geq 2p$}
\\ 0& \mbox{ otherwise}\ema\right.
\eqend
where $\Gamma$ is a grading operator on $\cH$ such that $\eps\Gamma =
-\Gamma\eps$.  For this also Stokes' theorem holds (due to the cyclicity of
trace), and under the homomorphism \Ref{c}, $\hat{\int}\hat \omega = \int
c(\hat\omega)$ for all $\hat\omega\in\hat\Omega_p'$.

As mentioned, it has been suggested that NCG should provide appropriate
mathematical tools for formulating and studying quantum field theory
models.  A universal Yang--Mills theory in this spirit has been proposed
and studied in \cite{R,FR}.  In these cases, the models were designed such
that a formulation solely in terms of notions from NCG was possible.  We
believe that it would be extremely interesting to extend such an approach
to Yang--Mills theories of the usual kind.

The theorem of the present paper allows such a formulation for {\em
topological} Yang--Mills field theories (whose actions involve only
integrals of de Rham forms and no Riemannian structure, e.g.\
Chern--Simons theories) by using the embedding of Yang--Mills field
configurations $A\in\Omega_d^{(1)}$ in $\hat\Omega_p^{(1)}$ as discussed
above\footnote{I am grateful to J.~Mickelsson for pointing this out to
me}.  This should provide the first step to implement such a program for this
class of models.  To extend it to other Yang--Mills gauge theories would
require a generalization of the Hodge--$\star$ operation to the
non--commutative setting.

\app
\bc
\section*{Appendix A: $\TraL{\ccr{a}{\hat X}}$}
\ec
In this appendix we prove eq. \Ref{aX}. Using the notation from the proof
of Lemma 3 (eq.\ \Ref{Ka} {\em ff}) we have
\[
\TraL{\ccr{a}{\hat X}} = \intF\dvp\intS\dvq \, \tra{K(a)(p,q)\tilde X(q-p) -
\tilde X(p-q)K(a)(q,p)}.
\]
Using cyclicity of ${\rm tr}$ and changing variables to $Q=(-/+)(p-q)$ in
the first/second term, we get
\eq
\label{surface}
\TraL{\ccr{a}{\hat X}} = \intF\dvp\intS\dvQ
\, \tra{(K(a)(p\, |-Q)-K(a)(p-Q|-Q))\tilde X(Q) }
\eqend
where we introduced
\[
K(a)(p\, |Q) \equiv K(a)(p,p-Q).
\]
As mentioned in the introduction, this integral of the same kind as the
ones from Feynman diagrams giving anomalies (i.e.\ it would be zero for
$\Lam\to\infty$ if one could shift the integration variables, see e.g.\
\cite{JJ}).

We now use Taylor's expansion
\[
K(a)(p\, |-Q) - K(a)(q-Q|-Q) =\sum_{i=1}^d Q_i\f{\partial}{\partial
p_i}K(a)(p\, |-Q) - \f{1}{2}\sum_{i,j=1}^d Q_iQ_j \f{\partial^2}{\partial
p_i\partial p_j}K(a)(p-t\, |-Q)
\]
where $t=t(p,Q)$ with $0\leq t_i\leq Q_i$. Thus
\[
\TraL{\ccr{a}{\hat X}} =\intF\dvp\intS\dvQ
\sum_{i=1}^d \tra{ \f{\partial K(a)(p\, |-Q)}{\partial
p_i}  Q_i \tilde X(Q) ) } +R_\Lam.
\]
Comparing \Ref{Ka} with \Ref{symb} it is easy to see that the first
term on the r.h.s. of this eq.\ is equal to the first term on the r.h.s.
of \Ref{aX}. Moreover, using \Ref{stokes} we get
\[
R_\Lam = -\f{1}{2}\intF\dvp\intS\dvQ\, \delta(\Lam-|p|)
\sum_{i,j=1}^d \tra{ \f{p_i}{|p|} \f{\partial K(a)(p-t\,|-Q)}{\partial
p_j}  Q_iQ_j \tilde X(Q) ) },
\]
hence $R_\Lam=\OO(\Lam^{-1})$ follows from $K(a)(p\, |-Q) = \OO(|p|^{-d+1})$
and
$\tilde X(Q) = \OO(|Q|^{-\infty})$.

\appende

\app

\bc
\section*{Appendix B: Volume of $S^{d-1}$}
\ec
To make this paper self--contained we give an elementary proof of
\eq
\label{omd}
\Om_d = \intS \dd\xi\, \delta(1-|\xi|) = \f{2\pi^{d/2}}{\Gamma(d/2)}
\eqend
(volume of $S^{d-1}$). We have
\nonueqa
\Om_{d+1} = \intS \dd{\xi}
\int_{\R} \dd{y}\, \delta\left(1-\sqrt{|\xi|^2+y^2}\right) =\\
\Om_d \, \f{1}{2}\int_{\R} \dd{\rho}\, |\rho|^{d-1}\int_{\R} \dd{y}\,
\delta\left(1-\sqrt{\rho^2 + y^2}\right) = \\
\Om_d  \, 2\int_0^{\pi/2} \dd{\phi} \, (\cos(\phi))^{d-1} = \Om_d
\, \f{\Gamma(1/2)\Gamma(d/2)}{\Gamma((d+1)/2)},
\nonueqaend
(in the second line we introduced spherical coordinates with $\rho=|\xi|$),
thus
\nonueqa
\Gamma\left(\f{d+1}{2}\right)\Om_{d+1} =\Gamma(\f{1}{2}) \,
\Gamma\left(\f{d}{2}\right)\Om_d =
\ldots =  \Gamma(\f{1}{2})^{d/2}\, \Gamma\left(\f{1}{2}\right) \Om_1
\nonueqaend
and \Ref{omd} follows with $\Om_1=2$ and $\Gamma(\f{1}{2})=\pi^{1/2}$.
\appende

\bc{\bf Acknowledgments}\ec
I am grateful to G.~Ferretti, H.~Grosse, and S.G.~Rajeev and especially
J.~Mickelsson for helpful discussions.  I would like to thank the Erwin
Schr\"odinger International Institute in Vienna for hospitality where part
of this work was done.  This work was supported in part by the
"\"Osterreichische Forschungsgemeinschaft" under contract Nr.\ 09\slash
0019.


\end{document}